# Fast buffet onset prediction and optimization method based on a pre-trained flowfield prediction model

Yunjia Yang,[*] Runze Li,[†] Yufei Zhang,[‡] and Haixin Chen.[§]

*Tsinghua University, Beijing, 100084, People's Republic of China*

The transonic buffet is a detrimental phenomenon occurs on supercritical airfoils and limits aircraft's operating envelope. Traditional methods for predicting buffet onset rely on multiple computational fluid dynamics simulations to assess a series of airfoil flowfields and then apply criteria to them, which is slow and hinders optimization efforts. This article introduces an innovative approach for rapid buffet onset prediction. A machine-learning flowfield prediction model is pre-trained on a large database and then deployed offline to replace simulations in the buffet prediction process for new airfoil designs. Unlike using a model to directly predict buffet onset, the proposed technique offers better visualization capabilities by providing users with intuitive flowfield outputs. It also demonstrates superior generalization ability, evidenced by a 32.5% reduction in average buffet onset prediction error on the testing dataset. The method is utilized to optimize the buffet performance of 11 distinct airfoils within and outside the training dataset. The optimization results are verified with simulations and proved to yield improved samples across all cases. It is affirmed the pre-trained flowfield prediction model can be applied to accelerate aerodynamic shape optimization, while further work still needs to raise its reliability for this safety-critical task.

[*] Ph. D. candidate, School of Aerospace Engineering, email: yangyj20@mails.tsinghua.edu.cn

[†] Postdoctoral research assistant, School of Aerospace Engineering, email: lirz16@tsinghua.org.cn

[‡] Associate professor, School of Aerospace Engineering, senior member AIAA, email: zhangyufei@tsinghua.edu.cn

[§] Professor, School of Aerospace Engineering, associate fellow AIAA, email: chenhaixin@tsinghua.edu.cn (Corresponding Author)

## Nomenclature

| | | |
|---|---|---|
| $C_L$ | = | Lift coefficient |
| $C_D$ | = | Drag coefficient |
| $C_M$ | = | Pitching moment coefficient |
| $C_f$ | = | Surface friction coefficient |
| $C_p$ | = | Pressure coefficient |
| Ma | = | Mach number |
| Re | = | Reynolds number |
| $u, l$ | = | Class shape transformation coefficients of upper and lower surface |
| $L/D$ | = | Lift-drag ratio |
| $(t/c)_{max}$ | = | Maximum relative thickness of an airfoil |
| $w$ | = | flowfield |
| $x$ | = | airfoil geometry |
| $z$ | = | Latent code |
| $\alpha$ | = | Angle of attack |
| $\Lambda$ | = | Swept angle of the wing |
| $\eta$ | = | Spanwise location of the sectional airfoil |

Subscript / Superscript:

| | | |
|---|---|---|
| buffet | = | Under buffet condition |
| c | = | Under cruise condition |
| inf | = | Freestream value |
| sep | = | Under separation condition |

# I. Introduction

Supercritical airfoils are adopted in the design of modern transportation aircraft due to their high efficiency and robustness over a wide range of operating conditions[1]. Under the cruise condition, there is a supersonic region on the upper surface of a supercritical airfoil near the leading edge, and normally, it is followed by a stable shock wave that transitions the supersonic region to the subsonic region.

As the angle of attack (AOA) of the airfoil or the flight speed of the aircraft increases, the supersonic region on the upper surface of the supercritical airfoil expands, and the shock wave strengthens. When the shock wave is intensified to a critical value, it will interfere with the boundary layer and generate self-excited oscillation[2]. This leads to an unsteady pressure distribution and oscillating aerodynamic loads, known as buffet loads. The angle of attack at which the buffet loads occur is referred to as the buffet onset.

The transonic buffet load leads to performance loss and safety issues for the aircraft, hence limiting the aircraft's operating envelope. Since the last century, extensive research has been carried out to better understand the buffet phenomenon, as well as to delay and mitigate its effects. The conclusions from the research mentioned above can be found in the reviews by Lee[2] and Giannelis et al.[3].

From the perspective of the aviation industry, the primary concerns regarding the transonic buffet are how to accurately predict and effectively raise the buffet onset[4]-[13]. Although the transonic buffet is an unsteady phenomenon, the cost of experimental studies or unsteady simulations is prohibitively high for design and optimization purposes. Therefore, the industry developed empirical formulas[9] and buffet onset criteria based on steady simulation results, such as the trailing edge pressure divergence criterion[4], the lift-curve-break criterion[6], and the moment coefficient slope criterion[7]. Nonetheless, these methods still require simulations under multiple operating conditions to obtain buffet onset. For example, the widely used lift-curve-break criterion predicts the buffet onset from the lift curve, which is the curve of lift coefficients with respect to the AOAs, of the airfoil. Normally, lift coefficients under 20 to 30 AOAs are needed to determine the lift curve[11], leading to a massive computational cost even when applying the fastest Reynolds-averaged Navier Stokes (RANS) method.

In recent years, the development of machine learning (ML) techniques has caused breakthroughs in the rapid prediction method of buffet onset[12]-[14]. Xu et al.[12] developed a neural network surrogate model to predict the buffet magnitude and drag coefficient at a given angle of attack and combined the model with a genetic algorithm to optimize buffet performance. Li et al.[13] proposed a convolutional neural network (CNN) model to directly predict the buffet-onset angle of attack, and applied it as a constraint during optimization.

However, the direct machine-learning-based surrogate models between airfoil geometry and buffet onset share two major drawbacks, interpretability and generalizability. During training and application, the model acts as a "black box", making users hard to judge the reliability of model's output[15]. It is also questionable whether the model can be applied to various optimization problems, as it is sensitive to the bias in training dataset, and may fail to evaluate new samples generated with the optimization algorithm.

In our study, the ML technique is applied in a new approach. Instead of establishing the direct surrogate model, only the time-consuming RANS simulations in traditional buffet onset criteria are replaced with a pre-trained flowfield prediction model. The model is trained in advance and offline applied to optimization and without retraining. When facing a new airfoil geometry, the flowfields can be quickly obtained with the model, and the buffet onset can be predicted with traditional criteria based on these flowfields.

Compared to using the direct surrogate model, the proposed approach has a better visualization and generalization ability by predicting the flowfields as intermediates. On one hand, the predicted flowfields give users an intuitive way to test the reliability of model output by judging whether they are aligned with prior physical knowledge. They also make it possible to apply multiple buffet onset criteria to the predicted flowfields, so that the predicted buffet onset can be cross-validated. On the other hand, as machine learning models are highly data-dependent, the predicted flowfields can act as a regularization to help the model avoid overfitting the training dataset.

The new buffet onset prediction approach is employed in a multi-objective genetic algorithm to verify its effectiveness. Airfoils with different geometry characteristics and operating conditions are used as baselines, and their buffet onset and the cruise lift-drag ratio are simultaneously optimized under engineering constraints to demonstrate model's performance.

## II. Machine-learning-based buffet onset prediction methodology

The current study focuses on predicting and optimizing the buffet onset of two-dimensional airfoils with steady flowfield data. Although the transonic buffet is inherently a three-dimensional and unsteady phenomenon, many previous studies on buffet onset criteria were conducted on two-dimensional and steady simulations, and proved practical to capture the buffet mechanism[16] and guide the aerodynamic design for engineering purposes[13]. Table 1 includes several buffet criteria based on two-dimensional steady simulations that are reported in the literature.

Table 1.    Some existing buffet onset criteria

| | Buffet criterion | Description | Literature |
|---|---|---|---|
| **Category I: Criterion based on flowfield features** | | | |
| 1 | Trailing edge pressure divergence | The pressure coefficient on the upper surface at 95% chord deviates from the linear regime when increasing AOA | Pearcy and Holder[4] Chung et al.[7] Petrocchi and Barakos[11] Golestani et al.[17] |
| 2 | Shock wave location | The shock wave reaches the maximum streamwise location of the upper surface when increasing AOA | Chung et al.[7] Golestani et al.[17] |
| 3 | Shock wave location v.s. maximum camber location | The shock location is aft of and sufficiently close to the airfoil maximum upper-surface camber location | Iovnovich et al.[16][18] |
| **Category II: Criterion based on aerodynamic forces** | | | |
| 4 | Lift curve break | The intersection of the lift curve and the shifted linear regime of the lift curve | Clark and Pelkman[6] Chung et al.[7] Kenway and Martins[10] Petrocchi and Barakos[11] Li et al.[13] Paladini et al.[19] |
| 5 | The slope of moment coefficient | The deviation point of the pitching moment coefficient from the linear regime | Chung et al.[7] Petrocchi and Barakos[11] |
| **Category III: Criterion based on global stableness** | | | |
| 6 | Unstable oscillatory eigenmode | One eigenmode of the global stability analysis turns unstable | Crouch et al. [8][20] Timme et al. [21][22] |

The existing buffet onset criteria in Table 1 can be divided into several categories. Criteria 1 to 3 are based on the evolution of the flowfield features and they are directly derived from the buffet mechanism. However, until now, none of the proposed mechanisms has been shown to completely explain the buffet phenomenon. Criteria 4 and 5 are based on the evolution of the aerodynamic coefficients with respect to AOA. They are highly empirical, but proved to be effective in many cases. In addition, their results can be affected by the CFD settings such as the turbulence model and the computational mesh. There are also a promising kind of criteria based on global stability analysis as labeled category III, but it requires additional unsteady computation for the disturbance equation.

The traditional CFD-based buffet onset prediction methodology can be concluded as shown in Fig. 1(a). All these criteria rely on a series of flowfields or aerodynamic performance under different AOAs to determine the critical point that fulfills the standards. The simulations of these numerous flowfields with the computational fluid dynamics (CFD) method are time-consuming and limit the criteria to be applied in the optimization where a large number of individuals need to be evaluated.

To accelerate buffet onset prediction, machine-learning techniques can be applied to build a surrogate model to bypass CFD simulations during optimization. The most straightforward way is to directly map the buffet onset with airfoil geometry[13] as shown in Fig. 1(b). The surrogate model is initially trained on a large database that contains various flowfields of different airfoil geometries and operating conditions; then, it can be deployed offline to different optimization problems with different baseline airfoils and different constraints.

In the present study, the surrogate model is built in another way to pursue better interpretability and generalizability. As shown in Fig. 1(c), only the CFD simulation is replaced with the prediction by a pre-trained machine-learning model, and the rest procedures remain the same as in the traditional buffet onset prediction methodology.

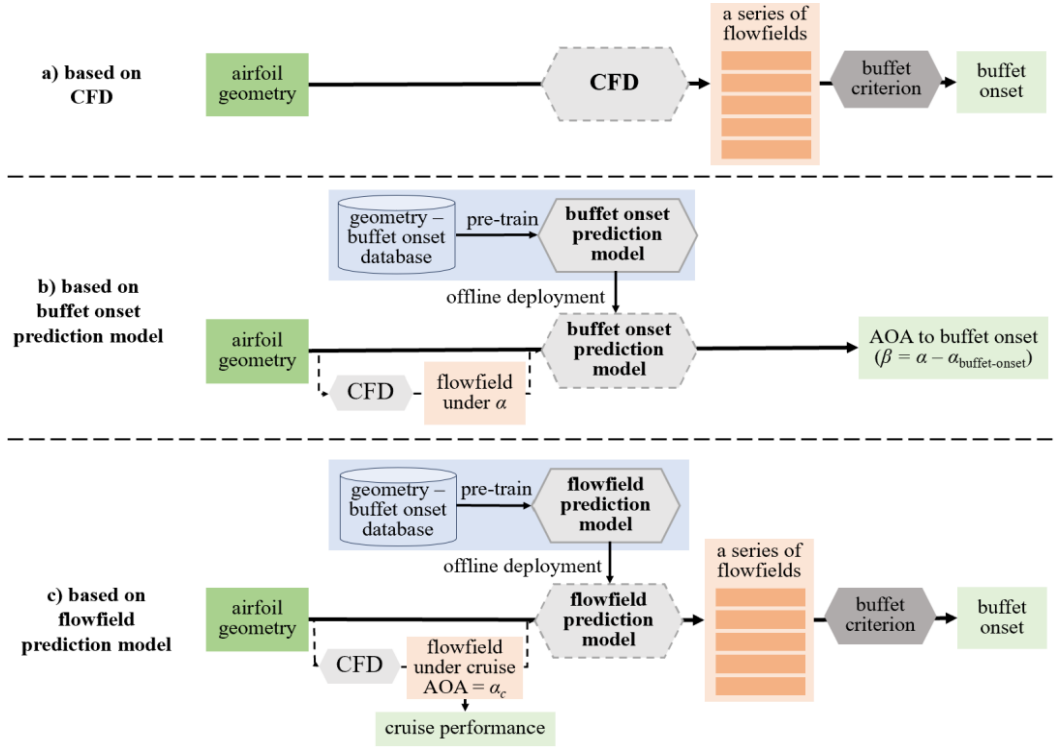

Fig. 1　CFD-based and model-based routines to predict buffet onset

As mentioned before, the proposed methodology can be applied with different buffet criteria. In the present paper, the most widely used lift-curve-break criterion is selected. As shown in Fig. 2, the lift-curve-break criterion identifies buffet onset at the intersection point of a straight line, which is formed by shifting the linear section of the lift curve 0.1 degrees to the right, and the lift curve itself.

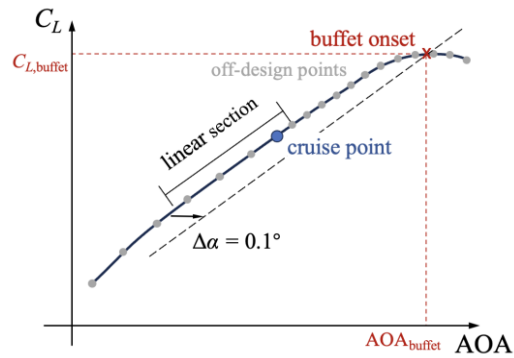

Fig. 2　Sketch of the lift-curve-break criterion

The traditional way to use lift-curve-break criterion is to obtain the lift curve by using CFD simulations under multiple AOAs. The lift coefficients can be integrated from the surface pressure and friction distributions, and then the lift curve can be fit with them. In the proposed machine-learning-based approach, a model is introduced to replace CFD and to quickly predict the surface distributions. Notably, not all CFD simulations are replaced with the model during buffet onset prediction. It is shown in several literature[13][23] that by introducing a prior flowfield as well as the airfoil geometry as the input of the machine-learning model, it will gain better prediction accuracy and generalization ability. Therefore, the surface pressure and friction distributions under the cruise operating condition are still calculated with the CFD simulation in the proposed buffet onset prediction routine, as shown with the dashed arrows in Fig. 1(c). Then, it is inputted together with airfoil geometry to the model for predicting the distributions under other off-design AOAs.

Once trained, the model-based method can predict the buffet onset of a new airfoil by calling the CFD simulation only once. The CFD simulation for the cruise flowfield usually takes approximately 2 minutes, and the inference time of the pre-trained model will take only another 2 seconds. Therefore, the time for evaluating the buffet onset of individuals is greatly reduced to the same as that of cruise-performance-only optimization.

### III. Flowfield prediction model and buffet criterion

In this section, two key links of the proposed prediction methodology, the prediction model and the buffet criterion, are discussed. For the prediction model, the prior-UNet (pUNet) model is proposed, and for the buffet criterion, the separation characteristics are added to the empirical lift-curve-break criterion.

#### A. Prior-UNet prediction model

With the rapid development of deep learning models, researchers have utilized several kinds of model architecture to predict flowfields, including multiple layer perceptron (MLP)[24], encoder-decode or autoencoder (AE)[25]-[29], generative adversarial network (GAN)[30][31], graph neural network (GNN)[32], and deep operator networks (DeepONet)[33]. Among them, the encoder-decoder network has a simple architecture and performs good accuracy, so it is selected as the backbone of the proposed prior-UNet model framework.

In an encoder-decoder model as shown in Fig. 3 (a), the encoder deals with the geometry input and extracts a low-dimensional representation of the geometry, and the decoder generates the flowfield with the representation and condition code. The pUNet's framework differs from it in two aspects, as displayed in Fig. 3 (b). First, the information in the CFD-simulated cruise flowfield is utilized as a prior input to the model, and the model only needs to predict the difference between the off-design flowfields and the cruise flowfield. This modification is denoted as the prior-based prediction strategy. Second, skip connections are added between the encoder and the decoder and are based on the UNet framework. This can protect the localization information during feature processing.

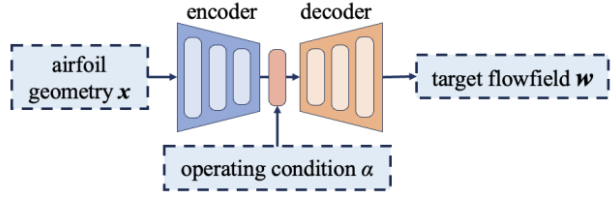

(a) Vanilla encoder-decoder framework for flowfield prediction

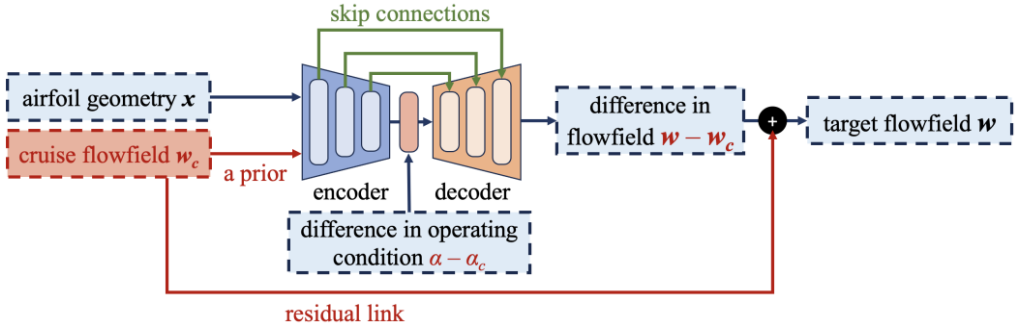

(b) pUNet framework for flowfield prediction

Fig. 3    Vanilla and pUNet frameworks

1. **Prior-based prediction strategy**

In the previous flowfield prediction models, the target flowfield is generated only from the airfoil geometry and the operating condition. The freedom of airfoil geometry is large, and the model has difficulty learning the mapping between various airfoil geometries to their flowfields. However, the evolution of the flowfield when increasing the airfoil's AOA obeys the same law for all airfoils; thus, the relationships between one airfoil's cruise flowfield and its off-design flowfields are similar for different airfoils. Consequently, the prior-based prediction strategy is proposed to introduce the cruise flowfield as a prior when predicting the target flowfield.

A theoretical proof is carried out to illustrate that introducing the cruise flowfield will help the prediction of the off-design flowfields. Suppose $w_c$ is the pressure and friction distributions at the cruising angle of attack, $\alpha_c$, for an airfoil $x$. The cruise distributions obey the governing equations of the flowfield $R = 0$. Thus, we have the following:

$$R(w_c, \alpha_c, x) = 0 \tag{1}$$

Let $w$ be the pressure and friction distributions at another angle of attack $\alpha$, which obeys the same governing equations of $R = 0$:

$$R(\boldsymbol{w}, \alpha, \boldsymbol{x}) = 0 \tag{2}$$

With Taylor expansion, we have the following:

$$(\boldsymbol{w} - \boldsymbol{w}_c) = \Delta \boldsymbol{w} \approx \frac{1}{(\partial R/\partial \boldsymbol{w})_c} \cdot \left(-\frac{\partial R}{\partial \alpha}\right)_c (\alpha - \alpha_c) \tag{3}$$

which can be reorganized as follows:

$$\boldsymbol{w} \approx \boldsymbol{w}_c + \mathcal{F}(\boldsymbol{w}_c, \alpha_c, \boldsymbol{x})(\alpha - \alpha_c) \tag{4}$$

where the symbol $\mathcal{F}$ represents a mapping function from the difference in $\alpha$ to the difference in the distribution; according to Equation (4), the mapping function relies on not only the geometry and operating parameters but also the cruise distribution. From the machine learning perspective, the model is used to fit the mapping function $\mathcal{F}$. By introducing $w_c$ into the prediction process, the model is theoretically more suitable for predicting the off-design flowfields and can be expected to have better precision and generalization ability.

Another advantage of the prior-based strategy is that the prior flowfield provides a baseline for the target flowfield. Previous research on residual learning has shown that the network can better fit a mapping relationship whose output is close to zero since the activation functions have the greatest gradient there[34]. A residual link is consequently deployed, as shown in Fig. 3 (b); thus, the model needs only to predict the difference between the target flowfield and the cruise flowfield.

2. UNet framework

After deciding the input and output of the model, the model framework is selected to realize the prediction in this section. As mentioned above, an encoder and a decoder are used to contact and reconstruct the flowfield, respectively, with multiple layers. The process is displayed in detail in Fig. 4. In the encoder, the kernels of the convolutional layers can be viewed as the extracted flowfield features that correspond to some statistical flowfield structures. The feature maps can be viewed as localization information of these flowfield features that represent how these flowfield structures are located in the space. As the number of convolutional layers increases, the abstraction level of the features rises, and the localization information about shallow features will be lost in a deeper network. The target flowfield shares similar flowfield features as the cruise flowfield; thus when the decoder attempts to upscale the target flowfield, the localization information of the features is important for its guidance.

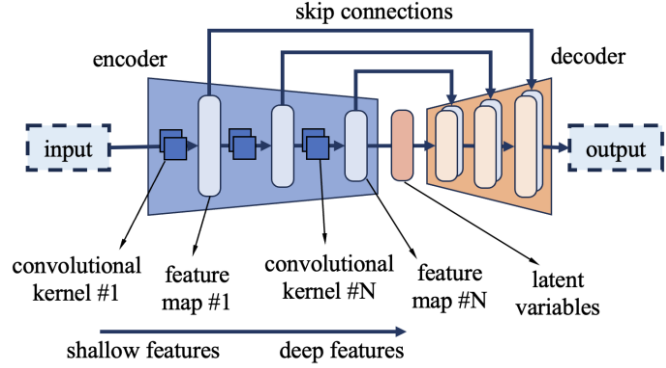

**Fig. 4**     **UNet framework**

The UNet framework from the ML community[35] is proposed to solve a similar need and has been introduced to the flowfield reconstruction tasks[36]-[38]. As shown in Fig. 4, the skip connections are built between the encoder and the decoder. The feature maps along the contracting path are saved and reinjected into the upscaling path by concatenating the information from the symmetric level of the contracting path. The skip connection can protect the localization information and reuse it during the upscaling process, thereby increasing the model performance. The UNet is initially designed to deal with images on a regular grid and has been extended to work better for CFD data on an irregular grid[38]. In the proposed pUNet model, the vanilla UNet is used and mesh grid coordinates are explicitly input to the model with surface distributions for simplicity.

### 3.    Setup of the model architecture

Fig. 5 describes the architecture of pUNet. The model consists of one encoder to contract the prior cruise distributions to a low-dimensional representation ($z$) and two identical decoders to generate the target pressure and friction distribution.

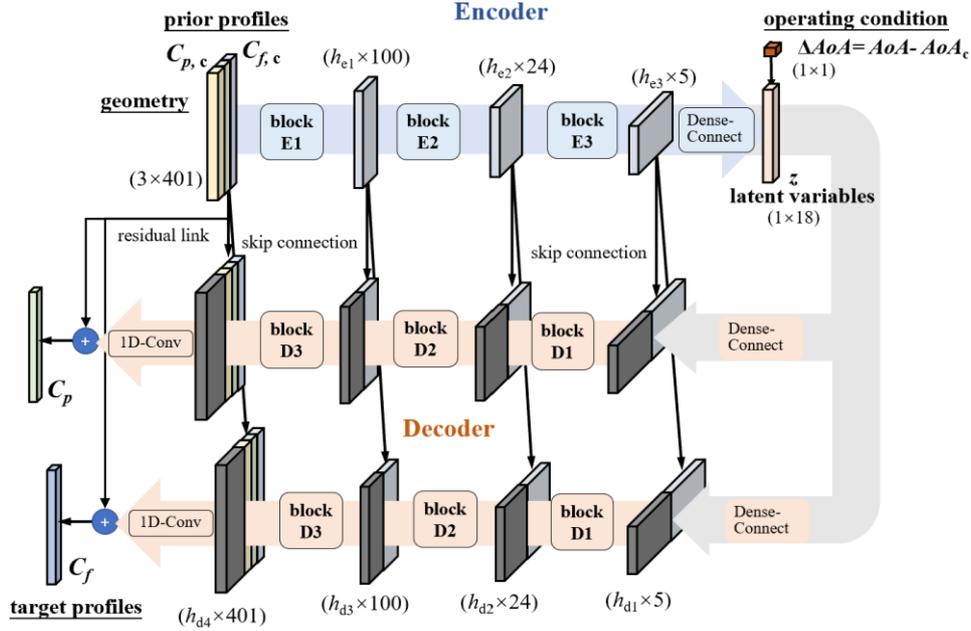

Fig. 5  The overall architecture of the pUNet model

The one-dimensional convolution layers are utilized to construct both the encoder and the decoder with the layer blocks, as shown in Fig. 6. The encoder comprises three blocks, each containing a 1D convolution layer with a kernel size ($k$) of 3 and a stride ($s$) of 2, an average pooling layer with the same kernel size and stride, and a LeakyReLU activation layer[39] with a slope of 0.2. The channel numbers after each block are expanded to $h_{e1}$, $h_{e2}$, and $h_{e3}$, which can be tuned for best performance. A densely connected layer then links the flattened output of the encoder to the latent vector, $z$.

Each decoder consists of three blocks that are similar to those in the encoder. The 1D convolutional layer is replaced with one linear interpolation layer for up-sampling the 1D feature map and a convolution layer with a stride of 1. The channel amounts of each feature map are $h_{d1}$ to $h_{d4}$. The network is concluded with another convolution layer that has a stride of 1 to compress the last feature map to one channel.

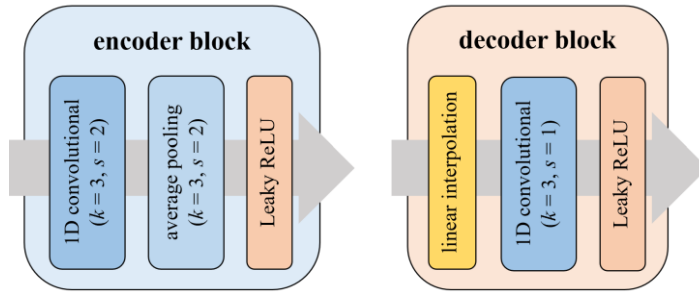

Fig. 6  Architecture of the encoder and decoder blocks in pUNet

The dimension of the latent variable is another tunable parameter in model framework. It is determined by several trial training around the degree of freedom of the problem and is fixed to 18 in the following sections. The network is implemented with the open-source package flowGen[1] based on PyTorch. There is also an illustration in flowGen's readme file on GitHub for the detailed architecture of the proposed model.

**B.  Enhanced buffet onset prediction criterion**

After flowfields are predicted with the pUNet model, the lift-curve-break criterion is applied to obtain the buffet onset. Although pUNet model is carefully designed to minimize its error, the error is inevitable and can be amplified by buffet onset criterion, resulting in an unreasonable buffet onset prediction given by the method. An enhanced lift-curve-break criterion is proposed to raise the robustness of buffet onset criterion and help avoiding failure of the criterion caused by error in flowfields. The AOA at which the initial separation emerges is set as the lower boundary of the buffet onset when using the traditional lift-curve-break method to predict the buffet onset. The physics behind this is that it is generally accepted that the buffet is strongly related to the self-excited interaction between the shock wave and the separation region. Specifically, the buffet can only occur after the separation zone emerges on the upper surface of the airfoil.

For the two-dimensional flow over an airfoil, the separation can be equated with the occurrence of a recirculation area, where the *x*-direction velocity at the first cell center above the wall is negative. In the present study, a common definition of 2D surface friction coefficient is used, the value of $C_f$ is proportional to the first layer velocity projection on the surface tangential direction, while its sign is given positive when the first layer x-velocity is positive[13][40]. Thus, the recirculation can be indicated with a negative $C_f$ value occurring on the upper surface, and can consequently identify the separation.

The detailed routine is shown below.

| **Algorithm 1** Enhanced lift-curve-break criterion |
| --- |
| > Given the lift curve $\{AOA_i, C_{L,i}\}_{i=1, 2, ..., N}$ |
| > Decide the initial start point of the linear section: $i_l \leftarrow 0$ |
| > Obtain the initial separation AOA: |
|         $AOA_{sep} \leftarrow$ the first $AOA_i$ that separation occurs on upper surface |

---

[1] https://github.com/YangYunjia/floGen

> Decide the endpoint of the linear section:
> $i_u \leftarrow$ the first $i$ that the $r^2$ score of the linear regression for $\{AOA_i, C_{L,i}\}_{i=l,...,u}$ exceed 0.9999
> Obtain linear section slope and intersection: $k, b \leftarrow$ linear regression for $\{AOA_i, C_{L,i}\}_{i=l,...,u}$
> Find the intersection of the lift curve and the shifted linear section:
> $AOA_{buffet} \leftarrow$ intersection between $y = k(AOA - 0.1) + b$ and the lift curve
> if $(AOA_{buffet} < AOA_{sep})$
>   Output $AOA_{sep}$
> else
>   Output $AOA_{buffet}$

## IV. Database and model training

A database with flowfields of diverse airfoil geometries and under different operating conditions is critical to train a general flowfield prediction model aiming to be applied to various optimization problems. In this section, the supercritical airfoil flowfield database is established, and the pUNet model is trained on the database.

### A. Supercritical airfoil flowfield database

The database contains the flowfield of 1217 airfoils under multiple operating conditions. The operating condition includes the freestream condition, namely, the freestream Mach number, $Ma_{inf}$, the freestream Reynolds number, $Re_{inf}$, and the freestream temperature, $T_{inf}$, as well as the lift coefficient, $C_L$. The AOA of the airfoil is determined with the $C_L$ under a specific freestream condition.

In the database, one cruise operating condition and approximately 20 off-design operating conditions are assigned for each airfoil. The cruise condition for each airfoil has different $Ma_{inf}$ and cruise lift coefficients $C_{L,c}$, while $Re_{inf}$ is specified as 20 million for all airfoils. The off-design conditions are chosen to have different AOAs and the same freestream conditions as the cruise condition of the airfoil. The database is also open-sourced on Github in the same repository as pUNet.

#### 1. Sampling of the airfoil geometry and operating conditions

There are two steps in the sampling process. First, the airfoil geometry and its cruise operating condition are sampled with the output space sampling (OSS)[41] method. Then, the off-design operating conditions are determined for each airfoil with the adaptive method.

The OSS aims to obtain geometric parameters with more abundant and diverse pressure distribution patterns under cruise operating conditions. The detailed method can be found in the reference[41]. The airfoil geometry is parameterized by the class shape transformation (CST)[42] method. The upper and lower sides of the airfoil are represented by ninth-order base functions of CST, and the coefficients for each base function are denoted as $(u_i, l_i)$ $i = 0,1,\ldots 9$ for the upper and lower surfaces. The maximum relative thickness, $(t/c)_{max}$, is also a free parameter when reconstructing the airfoil with CST base functions. Hence, 21 parameters are involved to describe an airfoil. Moreover, the cruise operating conditions, $Ma_{inf}$, and $C_{L,c}$, are also assigned to each airfoil during OSS to obtain the cruise pressure distribution pattern. The sampling ranges of the parameters in this study are listed in Table 2.

Table 2.   Sampling ranges of the parameters in the output space sampling

| parameters | description | upper boundary | lower boundary |
| --- | --- | --- | --- |
| $u_0, u_1, \ldots, u_9$ | upper surface CST parameters | -2.0 | 2.0 |
| $l_0, l_1, \ldots, l_9$ | lower surface CST parameters | -2.0 | 2.0 |
| $(t/c)_{max}$ | maximum relative thickness | 0.09 | 0.13 |
| $Ma_{inf}$ | cruise freestream Mach number | 0.71 | 0.76 |
| $C_{L,c}$ | cruise lift coefficient | 0.60 | 0.90 |

The airfoil geometries sampled with OSS and their pressure coefficient ($C_p$) and friction coefficient ($C_f$) distributions under the cruise condition are depicted in Fig. 7.

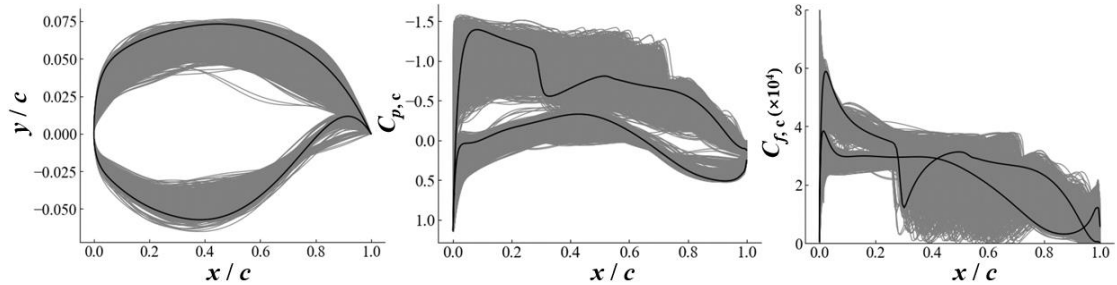

**Fig. 7**   Airfoil geometry, cruise pressure, and friction coefficient distributions in the database

The evolution of the airfoil's flowfield with increasing AOA is used in the selected lift-curve-break criterion; thus, several off-design conditions with different AOAs other than the cruise AOA are assigned for calculation.

An adaptive method is applied to each airfoil to obtain a series of AOAs. First, simulations are carried out on a coarse grid of AOAs from the range of -3.0° to 4.0° with an interval of 0.5°, and a predicted buffet onset is obtained with the lift-curve-break criterion on the results. Then, more AOAs are selected around the predicted buffet onset with an interval of 0.2. In total, 16 to 23 AOAs are selected, and the kernel density prediction of the AOAs for all airfoils with a bandwidth of 0.5 is depicted in Fig.

8.

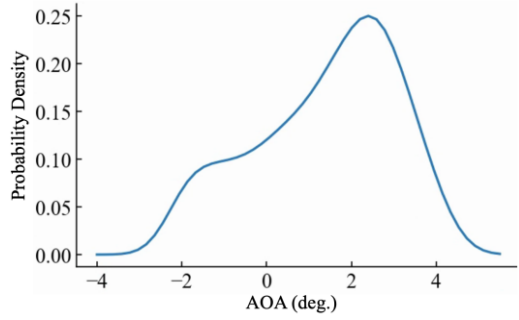

Fig. 8　　Kernel density prediction of all the AOAs in the database

2. **CFD methods**

The flowfields of the airfoils under cruise and off-design operating conditions are calculated with the open-sourced RANS solver CFL3D[40]. The solver is based on the finite volume method, where MUSCL scheme, ROE scheme, and Gauss-Seidel algorithm are adopted for flow variable reconstruction, spatial discretization, and time advance. The shear stress transport (SST) model is adopted for turbulence modeling. The structured C-type grid is used in the calculation and can be automatically generated using an in-house built code. The far-field location is 80 chords away from the airfoil. The grid contains 300 cells on the airfoil surface, 40 cells in the wake, and 80 cells in the far-field direction. The non-dimensional height of the first mesh layer $y^+$ is set to be below 1.

It is worth mentioning that computation mesh, turbulence modeling and other CFD settings may have a significant influence on the buffet onset predicted with the flowfields, so they are selected carefully with experience in our previous studies[23] and other literature[14]. The pressure coefficient distributions of a typical supercritical airfoil OAT15A are computed with different grid sizes and compared with the experimental results[43], as shown in Fig. 9. The three grids have 201, 301, and 401 grid points on the airfoil surface.

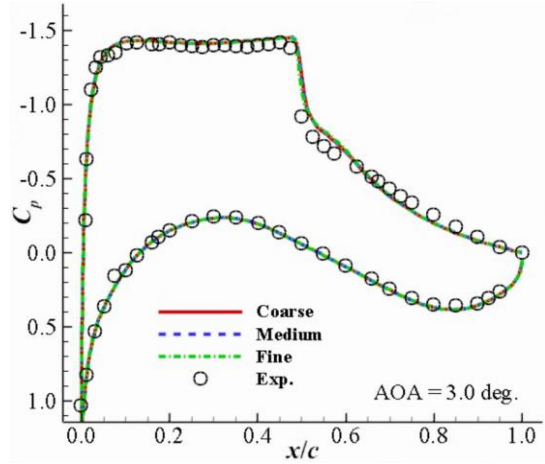

**Fig. 9   Pressure coefficient distributions of the RAE2822 airfoil under Ma=0.73, Re=3×10⁶ and AOA=3.0°**

3.   **Extraction and formatting of the surface distribution**

To reduce the difficulty of model training, the pressure coefficient distribution, $C_p$, and the friction coefficient distribution $C_f$, on the airfoil surface are selected to represent the entire airfoil flowfield in our study. Previous research has concluded that the transonic buffet is strongly related to the shock wave-boundary layer interaction[3],[13]. The shock wave features and patterns can be depicted with the surface pressure distributions, and the flow features in the boundary layer, such as the separations, can be captured with the surface friction distributions.

When establishing the database, the pressure and friction distributions are extracted from the simulated flowfield. The pressure can be directly obtained from the first layer on the airfoil surface, while the friction can be calculated from the gradient of the tangent velocity values between the first and second layers. Since $C_f$ has a smaller scale than $C_p$ by definition, the surface $C_f$ distribution is multiplied by 100 to make it a similar quantity as the $C_p$ distribution for the model to better learn both.

The geometry coordinates, surface pressure, and friction distributions of all samples are then formatted by interpolation to a series of specially designed $x$ positions. They distribute more densely near the leading and trailing edge to better capture the dramatic change in surface pressure and friction distributions. On the lower surface, the distribution is formatted from the trailing edge to the leading edge on 201 positions, and reversely on the 201 positions on the upper surface. The formatting process is illustrated in Fig. 10.

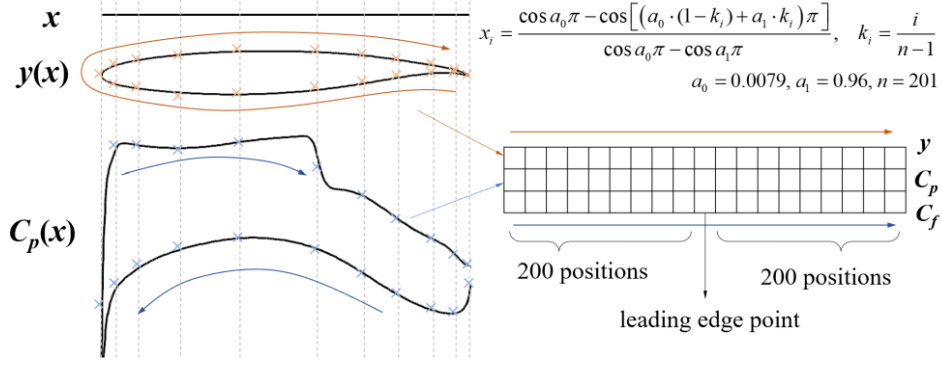

Fig. 10    Diagrammatic sketch of the data formatting process

### B. Model training

Before training, the flowfield database is randomly divided into a training set and a testing set. The flowfields from 1000 of the 1217 airfoils are selected to train the model, which gives a total of 21105 training samples. The remaining flowfields are only used for testing.

The loss function is selected to be the mean square error (MSE) between the model-predicted and CFD-simulated pressure and friction coefficient distributions. During training, a similar process as in Ref.[23] is adopted. A batch size of 16 is applied, and the optimizer is the Adam algorithm. The warmup strategy is employed to increase the learning rate from $5 \times 10^{-5}$ in the first 20 epochs and then reduced by an exponential function with a base of 0.95.

The training process is run three times to cross-validate the model. In each run, 10% of the samples are randomly selected from the training dataset for validation, and the model is trained on the remaining samples with a random initialization of weights and biases. The losses on the validation samples are monitored to avoid overfitting, and the maximum running epoch is set to 300. The training history plots can also be found in Appendix A.

## V.    Performance of the pUNet-based buffet onset prediction method

### A.    Prediction accuracy

As mentioned in the introduction, one of the advantages of using a flowfield prediction model in the buffet onset prediction process is that the model's output is more visible and intuitive. By evaluating the predicted surface pressure and friction coefficient distribution, it is possible for designers not only to judge the reliability of the result, but also to delve into the mechanism during the optimization. Therefore, the interpretability of the machine-learning-based model can be improved.

In this section, the pUNet's outputs are evaluated to test whether they can provide such information to designers. Surface pressure and friction coefficient distribution is the most intuitive representation of an airfoil, and are familiar to the designers. Two airfoils are randomly selected from the training dataset and one from the testing dataset, and their pUNet-predicted and CFD-simulated $C_p$ and $C_f$ surface distributions under several off-design operating conditions are depicted in Fig. 11. The ground truth CFD results are shown as black lines, while the distributions predicted with pUNet are depicted as red dashed-dots line.

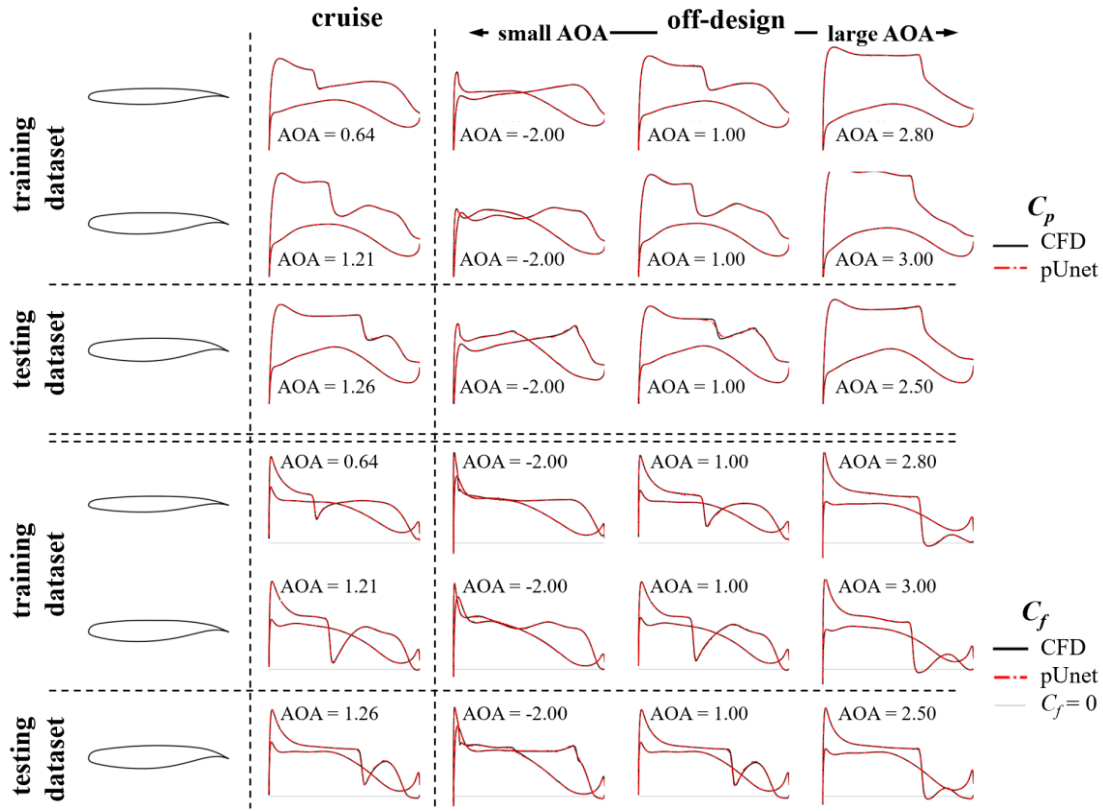

**Fig. 11**  $C_p$ and $C_f$ distributions predicted with the pUNet model with best training loss

Generally, pUNet can predict accurate distributions both on training and testing datasets, where the trend and turns are well captured. It can be seen from the figure that the error of the airfoil from the testing dataset is slightly larger but still acceptable. The pointwise mean square error (MSE) of the $C_p$ and $C_f$ distributions are shown in the left two columns of Table 3.

The flow structures such as the shock wave and separation are crucial information contained in the surface distribution to help designers understand the physics and evaluate the airfoil. They are also highly related to the occurrence of buffet phenomenon. Here, two flow features are selected to illustrate pUNet's ability on predicting flow structures. The first feature is the shock wave location ($X_{sw}$), which can be located from surface pressure coefficient distribution with the method proposed in Li's literature[15]. The second one is the separation occurrence on airfoil's upper surface. It can be obtained from surface friction coefficient distribution as mentioned in Sec. III. B. The errors of pUNet-predicted flow features are shown in middle two columns of Table 3. The shock wave location is measured with averaged absolute error, and the separation occurrence is measured by the percentage that the pUNet correctly predicted there is a separation or not. The pUNet achieves a shock wave location prediction error below 0.02 chord, and a separation prediction accuracy of 97%, indicating its results can be trusted for investigating physical mechanisms.

The averaged absolute prediction errors of aerodynamic coefficients are demonstrated in right two columns of Table 3. The coefficients are obtained by integrating surface pressure and friction distributions. The pUNet obtains a prediction error of the lift and drag coefficients of 0.0021 and 0.00025, respectively, which are reasonable for further application to buffet onset prediction.

Table 3. Performance of the pUNet model with best training loss

| dataset | surface distribution | | flow feature | | aerodyn. coefficient | |
|---|---|---|---|---|---|---|
| | $C_p$ | $C_f$ | $X_{sw}$ | $I_{sep}$ | $C_L$ | $C_D$ |
| training | 0.0055 | 0.000022 | 0.016 | 98% | 0.0021 | 0.00016 |
| testing | 0.0088 | 0.000043 | 0.019 | 97% | 0.0031 | 0.00025 |

The pUNet-predicted and CFD-simulated lift curves of the three airfoils in Fig. 11 are also shown in Fig. 12. The CFD-simulated lift curves are shown in black lines and circles, and the pUNet-predicted ones are in red lines and squares. The shifted linear sections by 0.1 degrees of the lift curves are shown in dash-dot lines, and the buffet onsets predicted by the two methods are shown in triangles at the intersection. The figure gives a good demonstration of the pUNet model's performance as the model-predicted and ground-truth CFD results are close to each other.

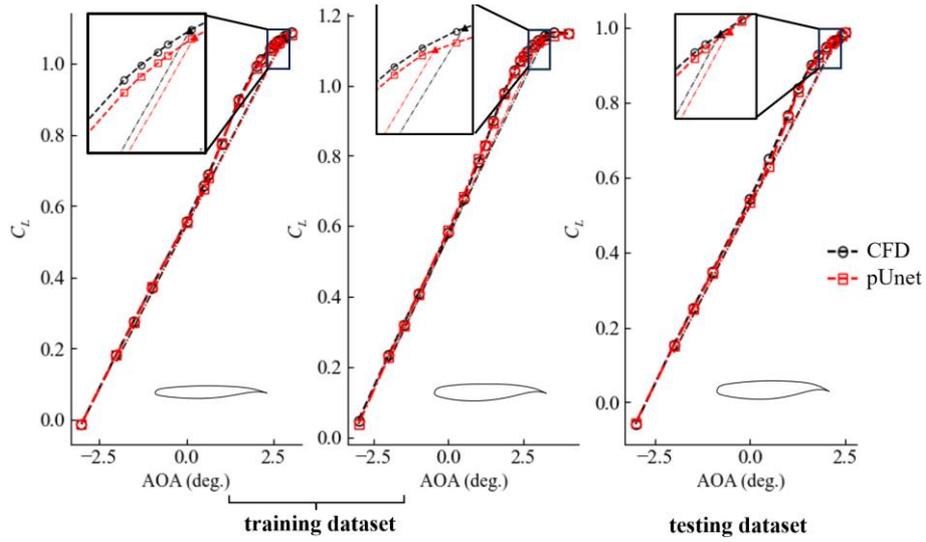

**Fig. 12   Lift curves predicted with CFD and the pUNet model**

## B.   Method generalizability

In addition to providing intuitive information to help improve the interpretability of buffet onset prediction method, the flowfield prediction models are also expected to have better generalization capabilities because their high-dimensional outputs will act as a regularization during training.

In this section, the proposed pUNet-based buffet onset prediction method is compared with two baseline models that take the buffet factor and the aerodynamic coefficient as output. The sketch diagrams of the model are shown in Fig. 13.

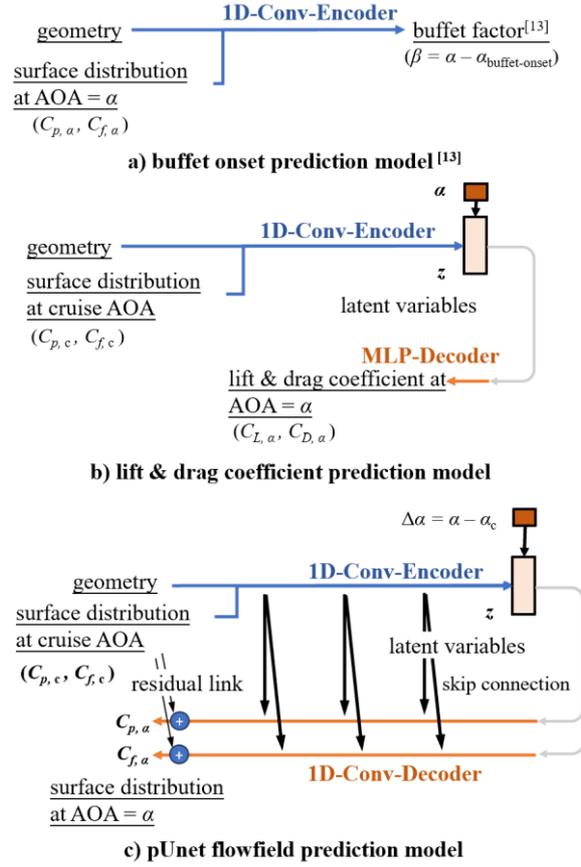

Fig. 13  Sketches of the architecture of the baseline and pUNet models

The baseline model A is a direct buffet onset prediction model that predicts the buffet factor (the difference between the current and buffet onset AOA) from airfoil geometry and current surface pressure and friction coefficient distribution. Its input and output are similar to the model in the literature[13], while the one-dimensional convolutional layers are adapted as its encoder's backbone to align with the proposed pUNet model. Model B is designed to predict lift and drag coefficients from the geometry and surface distributions at cruise AOA. The input and encoder part of baseline model B is the same as pUNet, while the decoder part of pUNet is replaced with a multiple layer perceptron (MLP) model to predict the coefficients. Both models A and B are set up to directly predict low-dimensional quantities (buffet onset, or coefficients) instead of the high-dimensional surface distribution. By comparing their predicting error with the error of the low-dimensional quantity obtained from postprocessing the surface distribution predicted by pUNet, the regularization effect of high-dimensional outputs can be demonstrated.

All the two baseline models are trained with the same training data of size 21105 and the same training settings as the pUNet model. The size of model trainable parameters has a large impact on model performance, and due to the pUNet model predicting the high-dimensional output, it should need a larger size of trainable parameters to reach its best performance. To eliminate the impact of model trainable parameters' size, networks with different numbers of hidden layers and hidden layer dimension sizes are implemented and trained. For each baseline model, the model architecture with the smallest prediction error on the testing dataset is used for the following comparison. The details of this process can be found in Appendix A.

The averaged absolute predicted error of aerodynamic coefficients of the pUNet and baseline B models are evaluated and compared in Table 4. For the pUNet model, the lift and drag coefficients are obtained by integrating the surface distributions that it predicts. The results show that the baseline B model achieves a better aerodynamic coefficient prediction error on the training samples, but the pUNet model can give smaller prediction errors on the testing samples. Although the advantage of the pUNet model is not significant, it is enough to prove that the prediction model with flow field as output can have better generalization performance in predicting aerodynamic coefficients.

Table 4. Performance of the pUNet models

|  |  | training samples | testing samples |
|---|---|---|---|
| dataset size |  | 18994 | 2111 |
| $C_L$ | pUNet | $20.10 \pm 0.29$ | **$32.50 \pm 0.30$** |
|  | baseline B | **$17.79 \pm 0.22$** | $32.89 \pm 0.72$ |
| $C_D$ | pUNet | $1.56 \pm 0.04$ | **$2.63 \pm 0.04$** |
|  | baseline B | **$1.29 \pm 0.02$** | $2.85 \pm 0.15$ |

The buffet onset prediction error is also evaluated. Here, the buffet onset obtained from CFD-simulated flowfields is seen as the ground truth for each airfoil. It is obtained by applying the enhanced lift-curve-break criterion to lift coefficients and separation features that are extracted from all the flowfields of the airfoil.

For baseline model A, the buffet onset for each airfoil is obtained by inputting the airfoil geometry and the cruise surface distributions into the model. Then the difference between cruise AOA and buffet onset AOA is predicted, thereby the buffet onset AOA is obtained. For baseline model B, the buffet onset is obtained by applying the traditional lift-curve-break criteria to a series of lift coefficients it predicts, since it does not predict any flow structure information. For pUNet, a series of lift coefficients are first integrated from the surface distributions it predicts, and both the tradition and enhanced lift-curve-break criteria are applied to obtain the buffet onset since the separation characteristics can also be extracted from the model's outputs.

The average and standard deviation of the buffet onset prediction errors are shown in Table 5. Again, the model that directly predicts the buffet onset achieves the best prediction error on the training dataset, and the average error on buffet onset AOA reaches 0.075 degrees, which matches the result given by Li et al. However, the baseline model A has the largest prediction error on the testing dataset. The prediction error of baseline model B is smaller than that of pUNet with traditional criteria on the training dataset, but on testing dataset, baseline model B only outperforms in buffet onset AOA. When applying the enhanced criterion to pUNet, the most accurate prediction on the testing dataset is obtained. Comparing to baseline model A, the averaged prediction error is reduced by 32.5% and 23.4% for buffet onset lift coefficient and AOA, respectively. This demonstrates that the better generalizability of the proposed pUNet-based buffet onset prediction methodology.

Table 5.  Performance of the baseline and pUNet models on buffet onset prediction

|  |  | training samples | testing samples |
|---|---|---|---|
|  | dataset size | 1000 | 217 |
| $C_{L,\text{buffet}}$ | pUNet (with enhanced criteria) | 0.0114±0.0006 | **0.0135±0.0008** |
|  | pUNet (with traditional criteria) | 0.0155±0.0009 | 0.0172±0.0034 |
|  | baseline B | 0.0138±0.0008 | 0.0186±0.0006 |
|  | baseline A | **0.0109±0.0002** | 0.0200±0.0004 |
| $AOA_{\text{buffet}}$ | pUNet (with enhanced criteria) | 0.0922±0.0039 | **0.1075±0.0031** |
|  | pUNet (with traditional criteria) | 0.1369±0.0067 | 0.1458±0.0197 |
|  | baseline B | 0.0995±0.0047 | 0.1250±0.0068 |
|  | baseline A | **0.0747±0.0014** | 0.1403±0.0018 |

In all, the proposed methodology shows better interpretability and generalizability compared to the direct buffet onset prediction model. However, the complexity of the proposed pUNet model is larger than the baselines, leading to a longer time to train and utilize the model, which will be discussed later. It may also require more experience and a longer process for trial-and-error to tune the model structure and hyperparameters for better model performance.

### VI. Multi-objective optimization for cruise and buffet performance

Though the proposed machine-learning-based buffet onset prediction method achieves reasonable prediction accuracy on training and testing datasets, its effectiveness in real optimization problems still needs verification.

In the present paper, the optimization problem is set to find the optimal airfoils with better cruise lift-drag ration and buffet onset given a baseline airfoil. The baseline airfoil is used as a starting point for optimization, and its thickness, cruise pitching moment, and other characteristics are required to be mostly maintained during optimization. To demonstrate the performance of the proposed method, airfoils both inside and outside the training dataset are selected as the baselines for optimization.

### A.   Optimization method

#### 1.   Optimization framework

The differential evolution (DE) algorithm[44] is an effective implementation of the genetic algorithm, and is a global, gradient-free, and stochastic optimization method. It finds the optimal sample by iteratively updating a sample population through operations that mimic biological evolution, such as mutation, crossover, and selection of individuals in the population. The differential here means that the mutation of the existing individuals is done by adding a differential vector between two randomly selected individuals to it.

In the present paper, the in-house program AeroOPT is used. It made several improvements to the origin DE algorithm and also combined it with local search on a respond surface to help generate new candidate individuals. The flow chart of the optimization is shown in Fig. 14, and a detailed algorithm can be found in the literature by Deng et al.[45] and Li et al.[46].

The proposed machine-learning-based buffet onset prediction method is utilized in the evaluation stage of the optimization, as also shown in Fig. 14. During the evaluation stage of each individual, its cruise flowfield is first obtained with the CFD simulation, and the cruise performance is calculated from the results. The surface distributions are also extracted from CFD results and inputted into the pre-trained pUNet model to predict a series of off-design surface distributions. Then, buffet performance can be obtained based on the criterion.

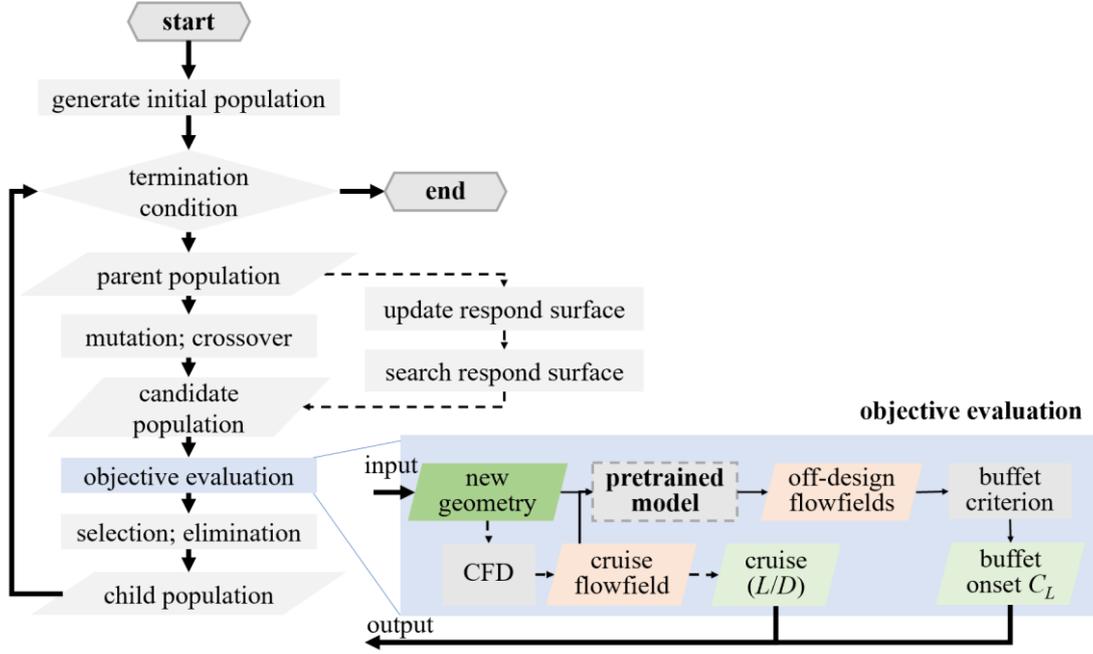

**Fig. 14    Overall routine for multi-objective optimization**

It is worth mentioning that the flowfield prediction model is a different model from the response surface in the optimization algorithm. In the present framework, all samples in optimization will not use the CFD-based method to obtain the buffet onset, so there will not be ground-truth buffet onset to update the pre-trained machine-learning model during optimization.

## 2.  Optimization setup

In optimization, the airfoil geometry is parameterized with 20 CST coefficient as well as the maximum relative thickness of the airfoil, which are the same as in Section IV when generating the supercritical airfoil database. Thus, there are a total of 21 design variables.

Table 6 concludes the objectives and constraints of the optimization. The objectives are to maximize the lift-drag ratio at the cruise condition and to maximize the buffet onset that is predicted with the pUNet model and enhanced lift-curve-break criterion. The cruise lift coefficient is constrained to be the same as the baseline, which is realized by the fixed lift coefficient scheme during CFD simulation; that is, the AOA is automatically adjusted to trace the required lift coefficient in the CFD iteration. The other constraints can be divided into two groups. Constraints 2 to 5 are used to limit the geometry and cruise characteristics (such as AOA and pitching moment) of the new airfoil similar to the baselines. The other three are empirical constraints to avoid unfeasible airfoils with bad off-design performance other than the transonic buffet.[46][47]

Table 6.  Objectives and constraints of the optimization

| | |
|---|---|
| Objectives | 1. Maximize the lift-drag ratio at cruise condition, $(L/D)_c$ <br> 2. Maximize the lift coefficient at buffet onset, $C_{L,\text{buffet}}$ |
| Constraints | 1. The cruise lift coefficient is the same as the baseline <br> 2. The maximum relative thickness, $(t/c)_{\max}$, is within ±2% of the baseline <br> 3. The thickness at 15% chord is not thinner than 95% of the baseline <br> 4. The AOA at cruise condition, $\text{AOA}_c$, is within ±0.5° of the baseline <br> 5. The pitching moment at cruise condition, $C_{M,c}$, is within ±10% of the baseline <br> 6. Only a single shock wave appears on the upper surface at cruise condition <br> 7. The shock wave position on the upper surface is behind 0.2 chord <br> 8. The wall Mach number in front of the shock wave not exceed 1.28 |

The initial populations are obtained by adding random bumps to the baseline airfoil and refitting the disturbed airfoil with CST functions. It helps to guarantee the modified airfoil is feasible and is similar to the baseline, so the property of the baselines can be maintained. A similar process can be found in reference[48]. The initial population size is 64, and the population size during optimization is 32. The optimization is ceased after 50 iterations.

**B. Baseline airfoils**

Three groups of airfoils both inside and outside the training dataset of the flowfield prediction model are selected as the baseline airfoil for optimization. The groups A and B are generated by scaling the two typical supercritical airfoils, the RAE2822[49] and the OAT15A[43], to a series of airfoils with different maximum relative thickness $((t/c)_{\max})$ from 0.08 to 0.14. The cruise Mach number and lift coefficient are determined by typical cruise conditions as shown in Table 7. The group C airfoils are the cross-sections of the common research wing (CRM) model[50] that are located at 26.38%, 54.87%, and 83.36% spanwise. Among them, the first airfoil is inside the kink, and the other two airfoils are beyond. The maximum relative thickness and cruise operating conditions of the three airfoils are converted with the swept theory[51] from those in the three-dimensional CRM case. The cruise Mach number of the wing is 0.85, and the sectional lift coefficients are obtained from reference[52]. The equivalent airfoil parameters for optimization are also summarized in Table 7.

Table 7.  Airfoil thickness and cruise operating condition for optimization

|  | Group A (OAT15A) | | | | Group B (RAE2822) | | | | Group C (CRM) | | |
| --- | --- | --- | --- | --- | --- | --- | --- | --- | --- | --- | --- |
|  | A1 | A2 | A3 | A4 | B1 | B2 | B3 | B4 | C1 | C2 | C3 |
| $(t/c)_{max}$ | 0.083 | 0.103 | 0.123 | 0.143 | 0.081 | 0.101 | 0.121 | 0.141 | 0.122 | 0.112 | 0.112 |
| Ma | | 0.730 | | | | 0.734 | | | 0.740 | 0.725 | 0.725 |
| $C_{L,c}$ | | 0.750 | | | | 0.700 | | | 0.821 | 0.861 | 0.738 |
| Re | | | | | 20 million | | | | | | |

The baseline airfoil geometries and their cruise pressure coefficient distributions are illustrated in Fig. 15. They are diverse in geometric features and cruise pressure distribution patterns. The airfoils have different maximum thicknesses and the thickest point is also located at different stations. Notice that the $(t/c)_{max}$ only ranges from 0.09 to 0.13 in the training dataset of the machine-learning model, thus the optimization A1, A4, B1, and B4 are beyond the training range of the model. By applying the proposed optimization method, its effectiveness and limitations can be demonstrated.

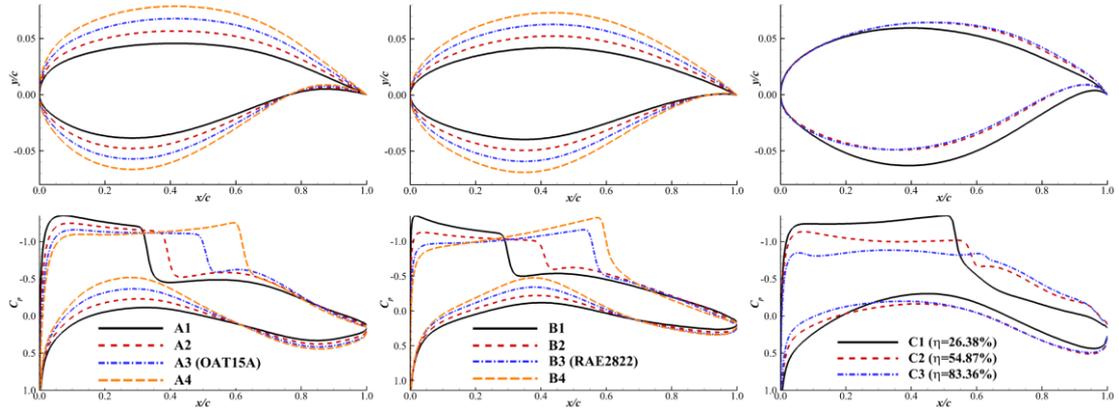

Fig. 15   Baseline airfoils' geometries and cruise surface pressure distributions

## C.   Optimization results

### 1.   Performance of model-based prediction method in optimizations

The optimization results for the test cases are displayed in Fig. 16, where the individuals are shown in the diagram based on the two objectives, where the x-axis is buffet onset lift coefficient and the y-axis is cruise lift-drag ratio. The baseline airfoil of each optimization is shown with blue circles, and the individuals in last population are shown with blue triangles. The triangles form a curve known as the Pareto frontier as shown in the blue solid line in diagrams. Notably, the $C_{L,\text{buffet}}$ for these individuals is predicted with the pUNet model, and the curve should be a model-based Pareto frontier.

To verify the results, the ground truth buffet onsets for the individuals in the initial and last population of each optimization case are calculated using the same criterion applied to CFD-simulated surface distributions. The CFD-calculated individuals' performance is shown with the red solid symbols, where the baseline ground truth is shown in circles, and the last population is shown in triangles. The latter also forms a Pareto frontier as shown in red dashed lines, which is denoted as the ground-truth Pareto frontier.

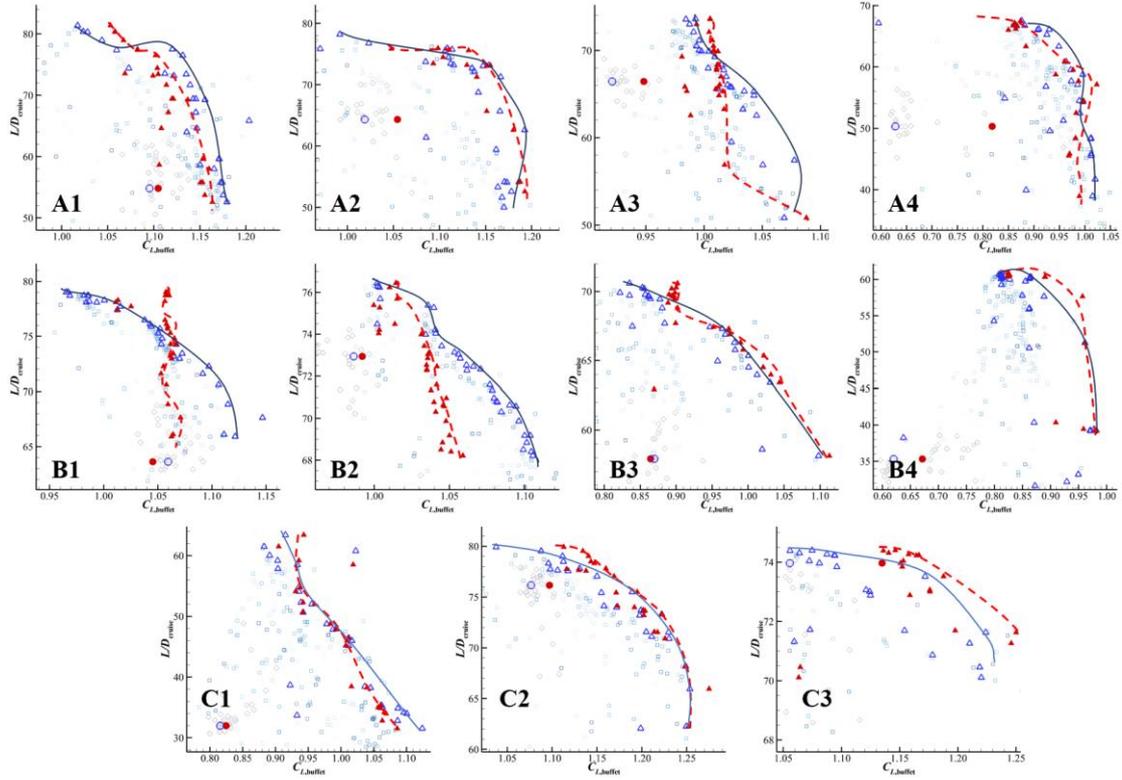

**Fig. 16    Optimization results on the diagram of the two objectives**

In Table 8, the model prediction errors of buffet onset lift coefficients for the initial and last population in every case are analyzed with the 90% confidential interval (CI90). A smaller length of 90% confidential interval (CI90$_{len}$) indicates the model has a better prediction accuracy on the case, and the middle point of 90% confidential interval (CI90$_{mid}$) shows whether the model over or under estimate the buffet onset. A negative value of CI90$_{mid}$ indicates the model underestimates the buffet onset, which will be a robust result.

**Table 8.    Statistics of model prediction errors on initial and last population for optimizations**

| case | | CI90$_{mid}$ | CI90$_{len}$ | CI90$_{mid}$ | CI90$_{len}$ |
|---|---|---|---|---|---|
| training dataset | | 0.000 | 0.003 | | |
| testing dataset | | – 0.004 | 0.010 | | |
| | | initial population | | last population | |
| OAT15A | A0* | – 0.008 | 0.007 | 0.012 | 0.017 |

|  | A1 | − 0.045 | 0.019 | − 0.005 | 0.021 |
|---|---|---|---|---|---|
|  | A2 | − 0.029 | 0.043 | 0.005 | 0.015 |
|  | A3* | − 0.178 | 0.149 | − 0.018 | 0.042 |
| RAE2822 | B0* | 0.012 | 0.010 | − 0.011 | 0.029 |
|  | B1 | − 0.020 | 0.044 | 0.030 | 0.012 |
|  | B2 | 0.001 | 0.013 | − 0.028 | 0.014 |
|  | B3* | − 0.031 | 0.031 | − 0.017 | 0.019 |
| CRM | C0 | − 0.002 | 0.006 | 0.000 | 0.017 |
|  | C1 | − 0.042 | 0.011 | − 0.018 | 0.013 |
|  | C2 | − 0.069 | 0.009 | − 0.034 | 0.052 |

* the baseline airfoil of this optimization case is beyond the training dataset

Both Fig. 16 and Table 8 demonstrate the model-predicted and ground-truth results are in good agreement. The model prediction errors on optimization samples are of the same magnitude as the testing dataset, except for cases A3 and C3. For the other three optimizations on the baseline airfoil outside training dataset (case A0, B0, and B3), the error is not higher than when optimizing within it, proving again the proposed method has a good generalization ability.

In addition, the model prediction errors in the last population of optimization are not consistently lower or higher than in the initial population. Plus, the model tends to under-predict buffet onset on the testing dataset and initial populations, but on the last populations, in several cases the model over-predict the results. They show that even if inherent bias exists in the prediction model, it has a limited impact on the optimization process, and the model-based optimization algorithm can still give reasonable and diverse results. However, they also show that the model prediction errors are highly dependent on problems, which calls for further research on model uncertainty quantification so that the model error can be evaluated and controlled during optimization.

2.  **Improvement in airfoil cruise and buffet performance**

The final gains from optimizations and the mechanism behind the optimized airfoils are analyzed in this section. A sample is selected from the last population of each optimization case with maximum buffet lift coefficient under the condition that the cruise lift-drag ratio is not lower than the baseline, and their performances are compared with baselines in Table 9. The buffet onset lift coefficient in the table is obtained with CFD data, and the percentages in the blanket next to the objective values are calculated by dividing the improvements by the baseline value. The results show new airfoils with larger buffet onsets and cruise lift-drag ratios are yielded in all eleven optimization cases.

Table 9. Comparison of the performances of optimized and baseline airfoils

| case | | baseline | | `sample with max. $C_{L,\text{buffet}}$ incr. under $(L/D)_c >$ baseline | |
|---|---|---|---|---|---|
| | | $(L/D)_c$ | $C_{L,\text{buffet}}$ | $(L/D)_c$ | $C_{L,\text{buffet}}$ |
| OAT15A | A0* | 1.10 | 54.82 | 1.16 (5.0%) | 58.01 (5.8%) |
| | A1 | 1.05 | 64.35 | 1.16 (9.6%) | 73.18 (13.7%) |
| | A2 | 0.95 | 66.45 | 1.01 (7.0%) | 67.01 (0.8%) |
| | A3* | 0.82 | 50.34 | 1.02 (25.1%) | 57.14 (13.5%) |
| RAE2822 | B0* | 1.05 | 63.62 | 1.07 (2.4%) | 67.64 (6.3%) |
| | B1 | 0.99 | 72.94 | 1.04 (4.5%) | 73.14 (0.3%) |
| | B2 | 0.86 | 57.93 | 1.11 (28.8%) | 58.14 (0.4%) |
| | B3* | 0.67 | 35.33 | 0.98 (46.4%) | 39.20 (11.0%) |
| CRM | C0 | 0.82 | 32.00 | 1.07 (29.6%) | 34.01 (6.3%) |
| | C1 | 1.10 | 76.17 | 1.17 (6.6%) | 77.05 (1.1%) |
| | C2 | 1.13 | 73.97 | 1.17 (2.8%) | 74.24 (0.4%) |

Then, the samples are analyzed in detail to show the mechanism of optimization. Due to the limitation in space, only the optimization results for the first two CRM sectional airfoils are analyzed in detail. The results for the other airfoils can be found in Appendix B.

Their CFD-simulated lift curves and surface pressure distributions under cruise and AOA larger than cruise are shown in Fig. 17. In the $C_L$ – AOA diagram on the left, the lift curve is shown with a dashed line, and the shifted linear section is shown with a dashed-dotted line. The initial separation and buffet onset are displayed with squares and triangles, respectively. The coordinate of the baseline and optimized airfoil is aligned at the cruise AOA for better comparison. The $C_p$ distributions of the airfoils under multiple AOAs are displayed on the right of each subfigure, and the cruise distribution is displayed with a background color. The baseline and optimized airfoil are also shown in each figure.

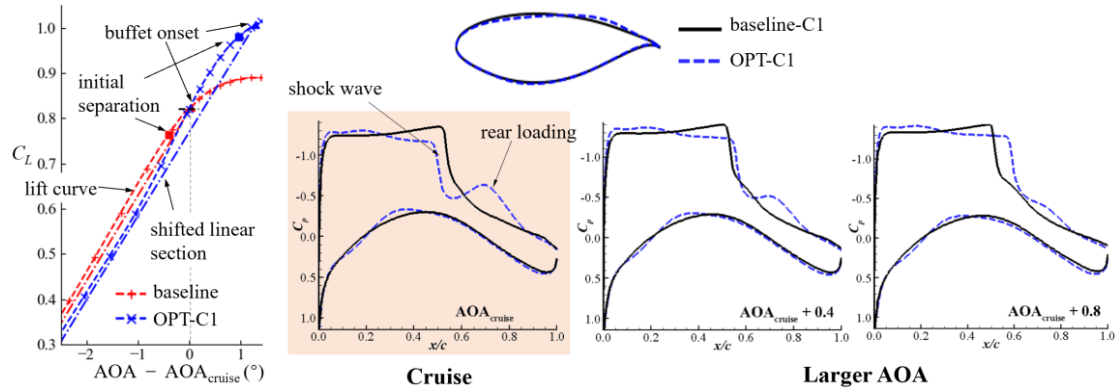

(a) Results for optimization C1

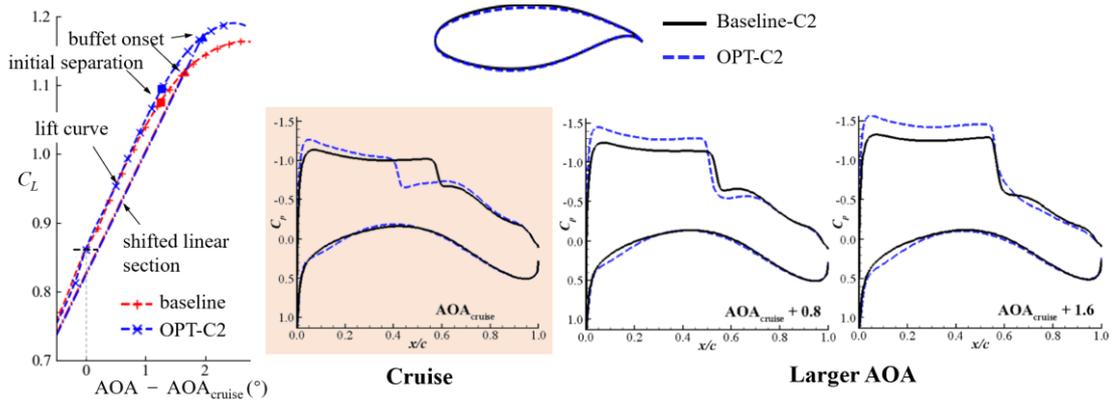

**(b) Results for optimization C2**

**Fig. 17** Lift coefficients and pressure distributions for the optimized airfoils

In both cases, the cruise shock wave locations on the optimized airfoil share a common trend that is moved upstream from the baseline. When increasing the AOAs of an airfoil, the shock waves will move toward the tailing edge. According to Iovnovich and Raveh[16], the buffet occurs when the shock wave reaches the maximum camber location of the airfoil. A more upstream cruise shock wave position increases the distance from the shock wave needed to reach the critical location, thereby delaying the buffet.

The shock wave strength is also important to separation and buffet behavior. For optimization C1, the cruise $C_L$ is high for baseline airfoil, resulting in a strong shock wave and separation on the upper surface under cruise conditions, and consequently, the buffet onset for baseline airfoil C1 occurs very close to the cruise AOA. After optimization, the shock wave strength is greatly reduced under cruise conditions and high AOA conditions, so the separation is delayed and the buffet performance is improved. For the optimization C2, the shock wave strength is medium, so after optimization, the shock wave strength remains the same.

These two reasons contribute to a higher buffet onset for all three optimization cases. Meanwhile, the maximum and 15% chord thicknesses, cruise AOA, and cruise pitching moment of the baseline airfoil are maintained due to the constraints during optimization. Nevertheless, the improvement in buffet onset has possible trade-offs in other performance. Since the cruise lift coefficient is fixed for the optimization, the area enclosed by the surface pressure coefficient distribution should be the same for the optimized and the baseline airfoil. A weaker and more upstream shock wave decreases the area enclosed by the distribution, so it needs to be compensated elsewhere on the diagram. For cases C1, the stronger rear loading occurs downstream of the shock wave, and for both optimized airfoils, the maximum Mach number on the lower surface increases. They help to maintain the fixed lift coefficient under cruise but may have negative impacts on other off-design performance like the drag-divergence. To overcome this issue, more constraints should be added to the optimization to help obtain the best optimization result.

3. Time cost

The costs of different methods are listed in Table 10. When using the CFD-based buffet onset prediction method, the only cost is to evaluate every individual with multiple CFD simulations. Assuming 16 off-design flowfields are used to locate the buffet onset, it will take 0.6 CPU hours to obtain the objective for one individual. This sums up a total of 943 CPU hours to conduct an optimization with 50 iterations and 32 individuals for each iteration.

For the model-based optimization method, the cost can be divided into two parts, time for building the model and the optimization based on the model. The time-consuming part is the former, and in the present work, about 25000 flowfield simulations are conducted to select valid samples for the dataset, which takes about 830 CPU hours. However, once the model is trained, the optimization can be done in only 55 CPU hours, plus 8 CPU hours to verify 20 samples on the model-predicted Pareto frontier with the CFD-based method. It is more important that the study above has proven that the model-based optimization can be applied to multiple problems even if the baseline airfoil is significantly different from the training ones. Therefore, the time invested in generating the database and training the model can be applied across various optimization problems, considerably accelerating the design process.

The cost of using pUNet is slightly larger than the direct buffet onset prediction model, but since they both need one CFD simulation for cruise flowfield, the total optimization time is similar for the two methods.

Table 10.  The time cost of different methods

|  | based on CFD | based on direct prediction model | based on pUNet |
|---|---|---|---|
| data preparation* | / | ~ 830 CPU hour | |
| model training** | / | 0.6 GPU hour | 1.5 GPU hour |
| get cruise performance for one sample | 2 CPU min. | 2 CPU min. | |
| get buffet performance for one sample | 32 CPU min. | < 1 GPU sec. | 1 GPU sec. |
| verify Pareto frontier* | / | 8 CPU hour | |

| total optimization time for one case | ~ 943 hour | ~ 63 hour |

*CFD simulation is conducted on a single CPU of 2.00 GHz
**ML model training and prediction are conducted on NVIDIA RTX 3060 GPU

## VII. Conclusions

In the present paper, a model-based buffet onset prediction methodology is proposed. The prediction process of buffet onset is accelerated by replacing the time-consuming CFD simulations with a pre-trained machine-learning flowfield prediction model called the pUNet. The new method is tested on datasets as well as real optimization problems. The contribution of the paper can be concluded as below:

1. The proposed buffet onset prediction methodology demonstrates better interpretability compared to the model that directly outputs buffet onset by outputting intuitive surface distributions that can be understood by designers. The proposed pUNet model can predict precise surface pressure and friction coefficient distribution with an averaged prediction error of 0.0088 and 0.000043, respectively. It can also accurately capture the crucial flow structure characteristics, such as having an accuracy of 97% on predicting whether separation occurs on the upper surface. They all contribute to providing designers with much information about the prediction results.

2. The proposed model also show better generalizability because its high-dimensional surface distribution output acts as a regularization to avoid overfitting during training. The buffet onset lift coefficients and AOAs obtained from pUNet's surface distribution output have a 32.5% and 23.4% smaller error on testing airfoils than the direct buffet onset prediction model, even if it has a larger error on the training dataset.

3. The proposed buffet onset prediction method is efficient and effective when utilized to optimize different baseline airfoils both inside and outside the training database. Combined with a differential evolution algorithm, it managed to yield samples with better buffet onset and cruise performance for 11 baseline airfoils tested in the present paper. The verification of the optimization results illustrates that model's inherent bias has little influence on optimization. The mechanism of better performance is analyzed for three CRM section airfoils and is attributed to a more upstream shock wave location on the upper surface of the optimized airfoils.

The current study affirms that the pre-trained model can be employed in the optimization of different baseline airfoils, which sheds light on the complex multipoint aerodynamic shape optimization problems that used to need numerous CFD simulations. Nevertheless, verification of the optimization results with the CFD-based method is compulsory to apply the machine-learning-based methodology to the safety-critical aerodynamic design process. It still needs further work to better quantify and control the uncertainty of the machine-learning model to make the design process more reliable.

## Acknowledgments

This work was supported by the National Natural Science Foundation of China No. 92052203, 12202243, 12372288, 12388101, and U23A2069.

Subsonic Wing with Continuous Morphing Trailing Edge." In *35th AIAA Applied Aerodynamics Conference*. Denver, Colorado: American Institute of Aeronautics and Astronautics. https://doi.org/10.2514/6.2017-4218.

**Appendix A.  Hyperparameters for models**

The selection of machine learning model's hyperparameters can highly affect model performance. The hyperparameters include the trainable parameter size that is determined by model architecture, and the model training parameters such as the mini-batch size, optimizer, and learning rate strategy. In the present paper, the hyperparameters are selected based on the experience in our previous study. The model architecture including the number of hidden layers and the hidden dimension of each layer are then tuned to achieve the best performance.

The model architecture and the prediction error of the pUNet model and the two baseline models on both the training and testing datasets are listed in Table 11. The architecture of the encoder and decoder is indicated by the hidden dimensions for every layer. The performance metrics are the same as in training for each model. That is, the MSE of surface pressure and friction coefficient distribution for pUNet, MSE of lift and drag coefficient for baseline model B, and MSE of buffet onset AOA and lift coefficient for baseline model A. The mean and standard deviation of prediction errors are estimated among the three cross-validation runs for each model.

Table 11.  Model architecture and prediction error of model architectures

| model | encoder architecture | decoder architecture | amount of parameters | error on the training dataset | error on the testing dataset |
|---|---|---|---|---|---|
| pUNet | 64, 128, 256 | 128, 128, 64, 32 | 591212 | $0.000642 \pm 0.000012$ | $0.002267 \pm 0.000129$ |
| | 64, 128, 256 | 256, 128, 64,64 | 739750 | $0.000541 \pm 0.000023$ | $0.002230 \pm 0.000056$ |
| | 64, 128, 256 | 256, 256, 128, 64 | 1354534 | $0.000382 \pm 0.000016$ | $0.002135 \pm 0.000018$ |
| | 64, 128, 256 | 512, 512, 256, 128 | 1873388 | $0.000449 \pm 0.000079$ | $0.002296 \pm 0.000166$ |
| | 64, 128, 256 | 512, 256, 256, 128 | 2266662 | $0.002846 \pm 0.003194$ | $0.004088 \pm 0.002546$ |
| baseline B | 64, 128, 256 | 32, 32 | 148724 | $0.000021 \pm 0.000002$ | $0.000049 \pm 0.000001$ |
| | 64, 128, 256 | 32, 64, 32 | 151860 | $0.000010 \pm 0.000000$ | $0.000040 \pm 0.000001$ |
| | 64, 128, 256 | 64, 128, 64 | 164948 | $0.000007 \pm 0.000000$ | $0.000034 \pm 0.000004$ |
| | 64, 128, 256 | 128, 256, 128 | 215700 | $0.000006 \pm 0.000001$ | $0.000034 \pm 0.000002$ |
| | 64, 128, 256 | 256, 512, 256 | 415508 | $0.000009 \pm 0.000000$ | $0.000033 \pm 0.000002$ |
| | 64, 128, 256 | 512, 512, 256 | 551700 | $0.000020 \pm 0.000008$ | $0.000044 \pm 0.000005$ |
| baseline A | 16, 32, 64 | / | 21105 | $0.001451 \pm 0.000065$ | $0.001906 \pm 0.000033$ |
| | 32, 64, 128 | / | 32514 | $0.000812 \pm 0.000054$ | $0.001575 \pm 0.000087$ |

| | | | | |
|---|---|---|---|---|
| 64, 128, 256 | / | 126466 | 0.000563 ± 0.000031 | 0.001390 ± 0.000180 |
| 128, 256, 512 | / | 498690 | 0.000474 ± 0.000023 | 0.001234 ± 0.000063 |
| 256, 512, 1024 | / | 1980418 | 0.000321 ± 0.000017 | 0.001069 ± 0.000017 |
| 512, 1024, 1024 | / | 4736002 | 0.000295 ± 0.000022 | 0.001075 ± 0.000044 |

Fig. 18 depicts the prediction errors with increasing the model complexity. The solid lines with solid symbols are the average prediction error on the training dataset of three training runs, and the dashed lines with hollow symbols are the errors on the testing dataset. The standard deviations of the average error among three training runs are depicted as the error bars in the figure.

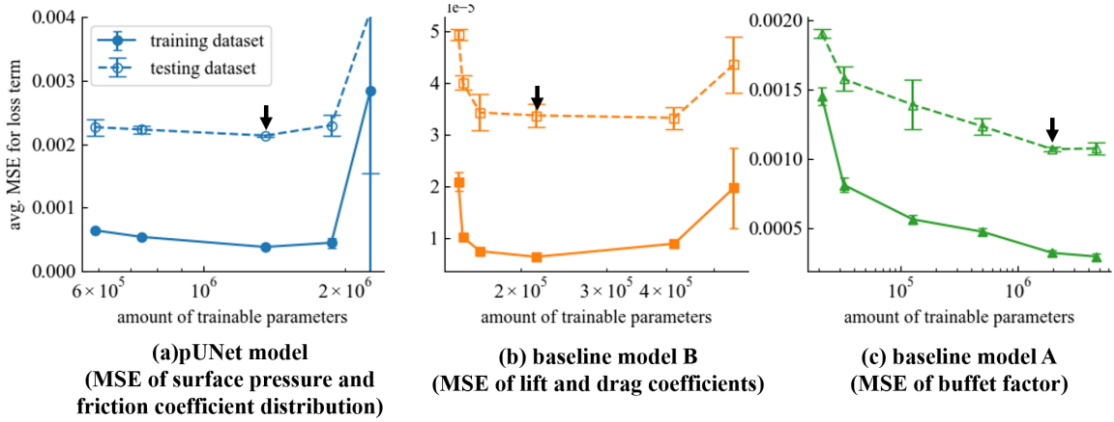

**Fig. 18  Training error of the three models with different amounts of parameter**

With increasing the trainable parameters in both models, the prediction errors on both training and testing datasets first decrease. During this period, the models appear to be underfitting the dataset. Then, the prediction error on the testing dataset begins to rise, while the error on the training dataset still decreases, which indicates the model starts to overfit the model. After the parameter amount reaches a certain value, the model becomes too large for current training method, leading the average and standard deviation of errors on both the training and testing datasets to rise.

The model sizes with the best prediction error on the test samples are marked with black arrows in Fig. 18. The pUNet and baseline A model achieve the best performance at a parameter amount of about 2 million, while the baseline B model has a smaller best size of about 0.2 million. The three best architectures are used to show the performance of the proposed method in Sections V and VI. Their loss histories during training are depicted in Fig. 19 indicating they are convergent.

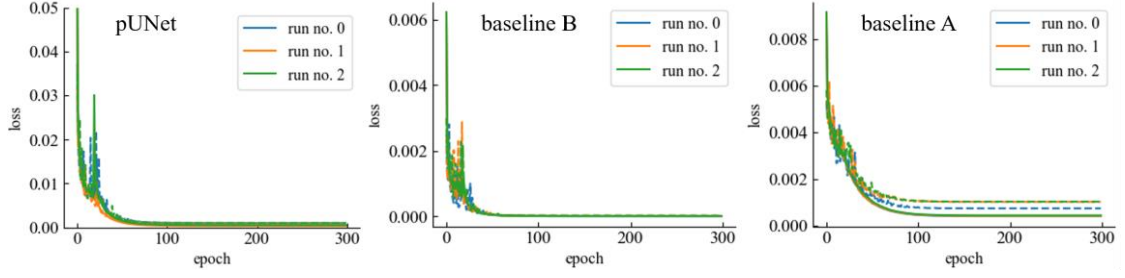

**Fig. 19  Training loss history of the three best architectures corresponding to pUNet and two baseline models (The loss on training samples for each run is shown in solid line, the loss on validation samples for each run)**

**Appendix B.  Performance of more optimized airfoils**

In Table 12, two more airfoils are selected from the Pareto frontier to demonstrate the optimization result and the mechanism behind it. The two additional samples are the individual with the maximum increase in buffet onset lift coefficient (marked as S1) and cruise lift-drag ratio (marked as S3). The individuals with the maximum gain in buffet onset under the condition that the cruise lift-drag ratio is not lower than the baseline (marked as S2), the same as in Table 9, are also listed. The performance in the table is obtained with the CFD-based buffet onset prediction method.

Table 12.  Performance of samples on the Pareto frontier

| case | | baseline | | S1 (max. $C_{L,\text{buffet}}$ incr.) | | S2 (max. $C_{L,\text{buffet}}$ incr. with $(L/D)_c$ cond.) | | S3 (max. $(L/D)_c$ incr.) | |
|---|---|---|---|---|---|---|---|---|---|
| | | $C_{L,\text{buffet}}$ (based on CFD) | | | | | | $(L/D)_c$ | |
| OAT15A | A0* | 1.10 | 54.82 | 1.16 (5.3%) | 52.60 (-4.0%) | 1.16 (5.0%) | 58.01 (5.8%) | 1.05 (-4.7%) | 81.40 (48.5%) |
| | A1 | 1.05 | 64.35 | 1.19 (13.1%) | 52.61 (-18.2%) | 1.16 (9.6%) | 73.18 (13.7%) | 1.10 (4.2%) | 75.96 (18.1%) |
| | A2 | 0.95 | 66.45 | 1.11 (17.2%) | 57.44 (-13.6%) | 1.01 (7.0%) | 67.01 (0.8%) | 1.00 (6.0%) | 73.63 (10.8%) |
| | A3* | 0.82 | 50.34 | 1.02 (25.1%) | 57.14 (13.5%) | same with S1 | | 0.87 (6.8%) | 67.54 (34.2%) |
| RAE2822 | B0* | 1.05 | 63.62 | 1.07 (2.4%) | 67.64 (6.3%) | same with S1 | | 1.06 (1.3%) | 79.03 (24.2%) |
| | B1 | 0.99 | 72.94 | 1.06 (6.7%) | 68.21 (-6.5%) | 1.04 (4.5%) | 73.14 (0.3%) | 1.01 (2.3%) | 76.44 (4.8%) |
| | B2 | 0.86 | 57.93 | 1.11 (28.8%) | 58.14 (0.4%) | same with S1 | | 0.90 (4.3%) | 70.59 (21.9%) |
| | B3* | 0.67 | 35.33 | 0.98 (46.4%) | 39.20 (11.0%) | same with S1 | | 0.82 (22.8%) | 60.83 (72.2%) |
| CRM | C0 | 0.82 | 32.00 | 1.09 (31.7%) | 31.56 (-1.4%) | 1.07 (29.6%) | 34.01 (6.3%) | 0.94 (14.4%) | 63.43 (98.2%) |
| | C1 | 1.10 | 76.17 | 1.27 (16.2%) | 65.97 (-13.4%) | 1.17 (6.6%) | 77.05 (1.1%) | 1.11 (1.5%) | 79.92 (4.9%) |
| | C2 | 1.13 | 73.97 | 1.27 (12.3%) | 51.61 (-30.2%) | 1.17 (2.8%) | 74.24 (0.4%) | 1.15 (1.0%) | 74.40 (0.6%) |

\* the baseline airfoil of this optimization case is beyond the training dataset

The airfoils and cruise surface pressure coefficient distributions of the samples above are shown in

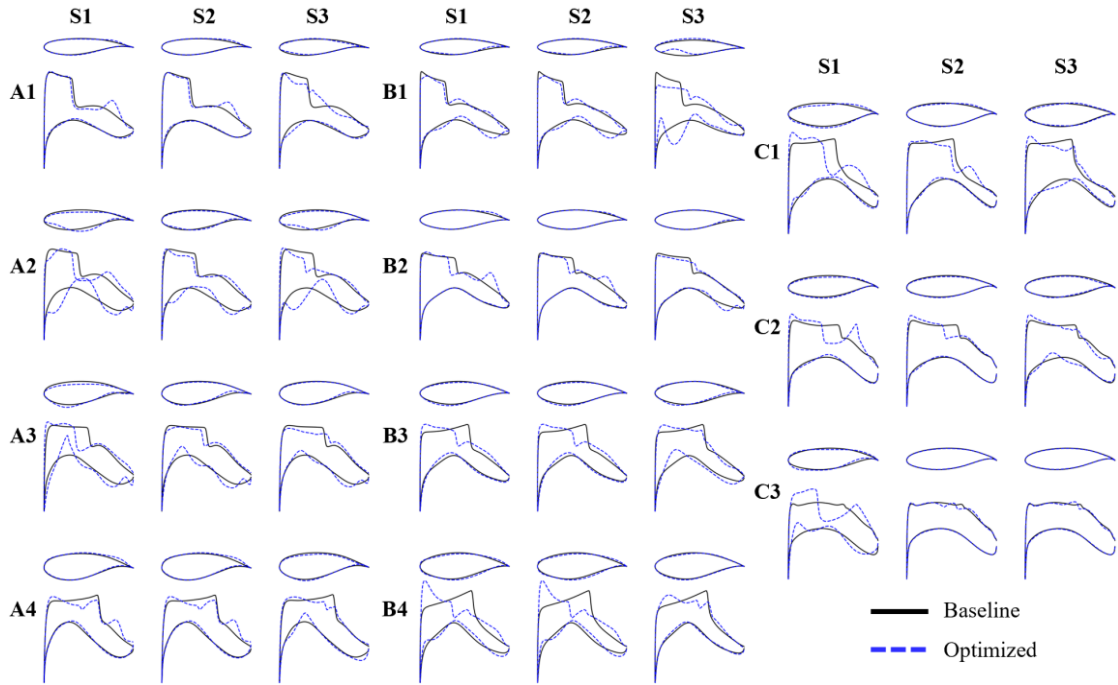

**Fig. 20  Geometry and cruise pressure distribution of optimized airfoils for cases**

There is a common trend among all optimization cases that the shock wave intensity is greatly reduced for the airfoil with maximum lift-drag ratio (column S3). It is not surprising because a portion of the drag for supercritical airfoils comes from the shock wave, and by alleviating the shock wave, the cruise lift-drag ratio can be improved. However, their buffet onset performance is not largely improved, and in some cases even lower than the baseline.

On the other hand, airfoils with the best buffet onset performance always have a shock wave located much upstream compared to the baseline. According to the analysis in Section VI, this can delay the AOA where the shock wave reaches the maximum camber location of airfoil, and thereby delay buffeting. However, these airfoils also have obvious drawbacks, such as strong flow acceleration on lower surface and strong rear loading. Their cruise lift-drag ratio is also lower than the baseline.

Different from the airfoils with extreme performance increment, the airfoils in the middle (S2) always have a better comprehensive performance, and may have a perspective to be applied in reality.

The results give a good illustration that maintaining a shock wave with proper intensity and position can lead to a better overall performance among cruise and off-design conditions. They also demonstrate that by evaluating samples in the last population of the optimization, airfoils with specific needs can be selected for further investigation.